\documentclass[a4paper,12pt]{article}

\usepackage[usenames]{color}
\usepackage{graphicx}
\usepackage{amsmath}
\usepackage{amssymb}
\usepackage{setspace}
\usepackage{xcolor} 

\begin{document}
\begin{titlepage}
\null
\begin{flushright}
July, 2017
\end{flushright}

\vskip 1.8cm
\begin{center}

  {\Large \bf ADHM Construction of (Anti-)Self-dual Instantons\\
\vspace{0.5cm}
in $4n$ Dimensions}

\vskip 1.8cm
\normalsize

{\bf  Koki Takesue\footnote{ktakesue(at)sci.kitasato-u.ac.jp}}

\vskip 0.5cm

  { \it
  Department of Physics \\
  Kitasato University \\
  Sagamihara 252-0373, Japan
  }

\vskip 2cm

\begin{abstract}
The ADHM construction is a very strong scheme to construct the instantons in four dimensions. 
We study an ADHM construction of instantons in $4n~(n\geq2)$ dimensions by generalizing this scheme.
The higher-dimensional ADHM construction generates the $4n$-dimensional (anti-)self-dual instantons which satisfy the (anti-)self-dual equation in $4n$ dimensions: $F(n)=\pm\ast_{4n}F(n)$.
Here $F(n)$ is the $n$th wedge products of the gauge field strength 2-form $F$.
We also show that our scheme reproduces the known $4n$-dimensional one-instantons and there are multi-instanton solutions of the 't Hooft type in the dilute instanton gas limit.
Moreover, we discuss a Harrington-Shepard type caloron in $4n$ dimensions and this monopole limit.
\end{abstract}
\end{center}
\end{titlepage}

\newpage
\tableofcontents
\section{Introduction}
Instantons in four dimensions are defined as solutions to the (anti-)self-dual equation $F=\pm\ast_4F$.
Here $F$ is the field strength $2$-form of the gauge field with a gauge group $G$ and the symbol $\ast_d$ is the Hodge dual operator in $d$-dimensional Euclidean space.
It is well known that instantons play important roles in the study of non-perturbative effects in gauge theories \cite{Belavin:1975fg,'tHooft:1976fv}.
Through the Bianchi identity, instantons are solutions to the equation of motion in the four-dimensional pure Yang-Mills theory.
The instantons are, moreover, characterized by the homotopy group $\pi_3(G)$, therefore we can classify these by the second Chern number $c_2=\frac{1}{8\pi^2}\int\text{Tr}[F\wedge F]$.
Of particular importance for the instantons is its systematic generation method of solutions, known as the Atiyah-Drinfeld-Hitchin-Manin (ADHM) construction \cite{Atiyah:1978ri}.
The ADHM construction algebraically constructs all the instantons in four dimensions, and the quaternion plays central roles on this algebraic side.

It is well known that the instantons are related to other lower dimensional solitons. For instance, a caloron is the soliton solution that we take a periodic direction in the instanton \cite{Harrington:1978ve}.
The dimensional reduction of the (anti-)self-dual equation to three dimensions leads to the Bogomol'nyi equation, and the BPS monopoles are defined as the solutions to this equation \cite{Prasad:1975kr}.
In the following, the monopoles mean the BPS monopoles.
There is the systematic construction, which is similar to the ADHM construction, of the monopoles and the calorons.
This construction is called the Nahm construction \cite{Nahm:1979yw}.
In three dimensions, there is an another soliton, known as Skyrmion, which is the solution of the static Skyrme model.
The Atiyah-Manton construction produces well-approximated solutions of the Skyrmions from the instantons \cite{Atiyah:1989dq}.

Naturally, we consider generalization of the four-dimensional instantons by generalizing the (anti-)self-dual equation to higher dimensions.
There are several kinds of ``instantons'' in higher dimensions, and these have been studied in various contexts.
One of the main types of instantons is sometimes called a secular type instanton \cite{Corrigan:1982th,Fubini:1985jm}.
The secular type instantons are the solutions to the linear equation $F_{\mu\nu}=\lambda T_{\mu\nu\rho\sigma}F^{\rho\sigma},\lambda\neq0, (\mu,\nu,\rho,\sigma=1,\dots,d)$. Here $d>4$ and the symbol $T_{\mu\nu\rho\sigma}$ is an anti-symmetric constant tensor
that respects subgroups of the SO$(d)$ Lorentz group.
However, the Chern numbers that are associated with the secular type instantons are not finite and quantized in general.
In this sense, the secular type instantons are not topological solitons.
An ADHM construction of the secular type instantons in $4n~(n = 1, 2, 3,\dots)$ dimensions has been studied in \cite{Corrigan:1984si}.

On the other hand, we can consider the another type equation which is the straightforward generalization of the four-dimensional (anti-)self-dual equation: $F(n)=\pm\ast_{4n} F(n)$.
Here $F(n)$ is the $n$th wedge products of the field strength 2-form $F$.
This equation is called the $4n$-dimensional (anti-)self-dual equation and solutions to this equation are called (anti-)self-dual instantons.
One of the most important characters of the (anti-)self-dual instantons is that these topological charges, which are defined by the $2n$-th Chern number $c_{2n}= \frac{1}{(2n)!(2\pi)^{2n}}\int\text{Tr}F(2n)$, are finite and quantized
when the homotopy group is non-trivial: $\pi_{4n-1}(G)\neq0$.
This type instanton was first studied by Tchrakian \cite{Tchrakian:1984gq},
\footnote{
The special case of $n=2$ was studied independently in \cite{Grossman:1984pi}.
}
and he constructed a spherical symmetry SO$(4n)$ instanton in $4n$ dimensions which is generalization of the four-dimensional Belavin-Polyakov-Schwartz-Tyupkin(BPST) instanton.
Furthermore, in $4n$ dimensions, an axially symmetric SO$(4n)$ one-instanton was presented explicitly in \cite{Chakrabarti:1985qj}.
This instanton is the analogy of the axially symmetric Witten solution in four dimensions. The existence of axially symmetric SO$(4n)$ multi-instantons has been proved analytically in \cite{Spruck:1997eb,Sibner:2003ee}.
However, in $4n~(n\geq2)$ dimensions, a SO$(4n)$ instanton of which symmetry less than axially symmetry does not exist \cite{Tchrakian:1990gc}.
From this fact, we propose the following question.
Can we construct higher-dimensional instantons of which other gauge groups and symmetries?
We will consider an approach of a higher-dimensional ADHM construction to elucidate this question.
In this paper, we treat only the case of which the base manifold is Euclidean space $\mathbb{R}^{4n}$.
Note that there are the (anti-)self-dual instantons on other base manifolds also, for instance, the case of complex projective space $\mathbb{C}P^m$ was discussed in \cite{Kihara:2008zg}.

In this paper, we study an ADHM construction of the $4n$-dimensional $(n\geq2)$ (anti-)self-dual instantons with the unitary gauge group U$(N)$.
The first non-trivial case $(n=2)$, the eight-dimensional ADHM construction, has been studied in \cite{Nakamula:2016srw}.
The $4n$-dimensional ADHM construction is generalization of the eight-dimensional one.
We will show that this is a general scheme to construct the (anti-)self-dual instantons and the known one-instantons in $4n$ dimensions can be reproduced from this scheme
\footnote{
Note that the gauge group of this reproduced one-instanton expand to the unitary group U$(2^{2n-1})$.
}.
Moreover, we will discuss higher-dimensional multi-instantons by introducing specific ADHM data which solve ADHM constraints, and we mention calorons and the monopole limit in higher dimensions.

The organization of this paper is as follows.
In the next section, we study the model that gives the (anti-)self-dual equation in $4n$ dimensions and review the known $4n$-dimensional one-instantons.
We introduce, moreover, an (anti-)self-dual tensor which is generalization of the 't Hooft symbol in four dimensions.
The (anti-)self-dual tensor is constructed from the complex Clifford algebra and plays the central roles of the higher-dimensional ADHM construction on the algebraic side.
In section 3, we study the ADHM construction of the U$(N)$ (anti-)self-dual instantons in $4n$ dimensions $(n\geq2)$.
We find that, in generally higher dimensions, there is an extra ADHM constraint in addition to the respected four-dimensional one.
This situation is same as the eight-dimensional case \cite{Nakamula:2016srw}.
One of the most interest things is that the ADHM construction in more higher than twelve dimensions does not require other new constraints.
In section 4, we consider some ADHM data in higher dimensions.
We first show that our construction precisely reproduces the well-known one-instanton profile of the solution.
Next, we consider a multi-instanton ADHM data which is generalization of the 't Hooft data in four dimensions and show that this data in the dilute instanton gas limit satisfies the ADHM constraints.
In section 5, we will discuss a $4n$-dimensional Harrington-Shepard type caloron and this monopole limit.
The last section includes the conclusion and discussions.
The explicit matrix representation of the complex Clifford algebra can be found in Appendix A.
We prove the existence of the inverse matrices of the ADHM constraints in Appendix B.

\section{(Anti-)self-dual instantons in $4n$ dimensions}
In this section, we study (anti-)self-dual instantons in $4n$-dimensional Euclidean space with the flat metric.
The $4n$-dimensional (anti-)self-dual equation is defined as the generalization of the usual four-dimensional (anti-)self-dual equation:
\begin{equation}
F(n) = \pm\ast_{4n}F(n),	\label{eq:4nd_ASD_eq_form}
\end{equation}
where $\ast_{4n}$ is the $4n$-dimensional Hodge dual operator, $F(n) = F\wedge\dots\wedge F$ ($n$ times) and $F=\frac{1}{2!}F_{\mu\nu}dx^{\mu}\wedge dx^{\nu}$ is the gauge field strength $2$-form of which component is defined by
$F_{\mu\nu} = \partial_{\mu}A_{\nu} - \partial_{\nu}A_{\mu} + [A_{\mu},A_{\nu}].$
Here $A_{\mu}$ is the anti-hermite gauge field ($A_{\mu}^{\dagger}=-A_{\mu}$) which takes value in a Lie algebra $\mathcal{G}$. The Lie algebra $\mathcal{G}$ is associated with the non-Abelian gauge group $G$ and the greek indices $\mu,\nu,\dots = 1,2,\dots,4n$ are the $4n$-dimensional Euclidean space indices.
The component expression of the (anti-)self-dual equation \eqref{eq:4nd_ASD_eq_form} is
\begin{equation}
F_{[\mu_1\mu_2}\dots F_{\mu_{2n-1}\mu_{2n}]} = \pm\frac{1}{(2n)!}\varepsilon_{\mu_1\mu_2\dots\mu_{2n-1}\mu_{2n}\nu_1\nu_2\dots\nu_{2n-1}\nu_{2n}}F_{\nu_1\nu_2}\dots F_{\nu_{2n-1}\nu_{2n}},	\label{eq:ASD_eq_components}
\end{equation}
where $\varepsilon_{\mu_1\mu_2\dots\mu_{2n-1}\mu_{2n}\nu_1\nu_2\dots\nu_{2n-1}\nu_{2n}}$ is the anti-symmetric tensor in $4n$ dimensions and the bracket $[\mu_1\mu_2\dots\mu_{2n}]$ means the anti-symmetrization of indices with the weight $1/(2n)!$.
The $4n$-dimensional (anti-)self-dual instantons are defined as the solutions to the $4n$-dimensional (anti-)self-dual equations \eqref{eq:ASD_eq_components}.

The action that gives the $4n$-dimensional (anti-)self-dual equation \eqref{eq:4nd_ASD_eq_form} is given by
\begin{equation}
S = (-1)^n\mathcal{N}_n\int\text{Tr}\left[ F(n)\wedge\ast_{4n}F(n) \right].	\label{eq:gYM_model}
\end{equation}
We call this action as the generalized Yang-Mills action.
Here $\mathcal{N}_n$ is the normalization constant in $4n$ dimensions which will be determined on the last in this section.
If we choose the hermite gauge field ($A_{\mu}^{\dagger}=A_{\mu}$) then the action coefficient signature is replaced $1$ instead of $(-1)^n$.
We easily show that the Bogomol'nyi completion  of the action is 
\begin{equation}
S = (-1)^n\frac{\mathcal{N}_n}{2}\int\text{Tr}\left[ \left( F(n)\mp\ast_{4n}F(n) \right)^2 \pm 2F(2n) \right] \geq \pm(-1)^n\mathcal{N}_n\int\text{Tr}F(2n),
\end{equation}
where we have defined
\begin{equation}
\left( F(n)\mp\ast_{4n}F(n) \right)^2 = \left( F(n)\mp\ast_{4n}F(n) \right)\wedge\ast_{4n}\left( F(n)\mp\ast_{4n}F(n) \right).
\end{equation}
The Bogomol'nyi bound of the action \eqref{eq:gYM_model} is saturated when the solutions satisfy the $4n$-dimensional (anti-)self-dual equation \eqref{eq:4nd_ASD_eq_form}.
Then the action is bounded from below by the $2n$-th Chern number $S=\pm(-1)^n\mathcal{N}_n\int\text{Tr}F(2n)$.

The $4n$-dimensional Belavin-Polyakov-Schwartz-Tyupkin(BPST) type instanton was discussed in \cite{Tchrakian:1984gq,Grossman:1984pi,OSe:1987fcz}.
We review this type instanton in the following.
The gauge field of the BPST type instanton is
\begin{equation}
A_{\mu}(x) = -\frac{1}{2}\frac{\tilde{x}^{\nu}}{\lambda^2 + \|\tilde{x}\|^2}\Sigma^{(\pm)}_{\mu\nu},	\label{eq:BPST_gauge_field}
\end{equation}
where we have defined $\tilde{x}^{\mu}=x^{\mu}-a^{\mu}$, $a^{\mu}\in\mathbb{R}$ is the position of the instanton, $\lambda\in\mathbb{R}$ is the instanton size and $\|\tilde{x}\|^2=(x^{\mu}-a^{\mu})(x_{\mu}-a_{\mu})$.
The symbol $\Sigma^{(\pm)}_{\mu\nu}$ is a $4n$-dimensional (anti-)self-dual tensor, and this is an analogy of the 't Hooft symbol in four dimensions.

The (anti-)self-dual tensor in $4n$ dimensions is given by
\begin{equation}
\Sigma^{(+)}_{\mu\nu} = e_{\mu}^{\dagger}e_{\nu} - e^{\dagger}_{\nu}e_{\mu},~~\Sigma^{(-)}_{\mu\nu} = e_{\mu}e^{\dagger}_{\nu}-e_{\nu}e^{\dagger}_{\mu},	\label{eq:4nd_ASD_tensor}
\end{equation}
with $e_i,e_i^{\dagger}$, which we call the (anti-)self-dual basis in $4n$ dimensions, are defined by
\begin{equation}
e_{\mu} = \delta_{\mu\#}\mathbf{1}_{2^{2n-1}} + \delta_{\mu i}\Gamma^{(-)}_i,~~e_{\mu}^{\dagger} = \delta_{\mu\#}\mathbf{1}_{2^{2n-1}} + \delta_{\mu i}\Gamma^{(+)}_i,~~~(i=1,\dots,4n-1,~~\#=4n),	\label{eq:4nd_ASD_basis}
\end{equation}
where $\Gamma^{(\pm)}_i$ are $2^{2n-1}\times2^{2n-1}$ matrices that satisfy the relation $\{ \Gamma^{(\pm)}_i,\Gamma^{(\pm)}_j \} = -2\delta_{ij}\mathbf{1}_{2^{2n-1}}$, and $\mathbf{1}_{2^{2n-1}}$ is the identity matrix.
The element $\Gamma^{(\pm)}_i$ is defined by $\Gamma_i^{(\pm)}=\frac{1}{2}(1\pm\omega)\Gamma_i$ and we choose $\Gamma_i^{(\pm)}$ that satisfies the relation: $\Gamma^{(+)}_i = -\Gamma^{(-)}_i$.
Here $\Gamma_i$ is the matrix representation of the $(4n-1)$-dimensional complex Clifford algebra: $\Gamma_i\in C\ell_{4n-1}(\mathbb{C})$, and $\omega$ is the chirality element which is defined by
\begin{equation}
\omega = (-1)^{n+1}\Gamma_1\Gamma_2\dots\Gamma_{4n-1}.
\end{equation}
The explicit matrix representation of the $(4n-1)$-dimensional complex Clifford algebras can be found in Appendix A.
The (anti-)self-dual basis $e_{\mu}$ is the generalization of the quaternion basis in four dimensions
and is normalized as $\text{Tr}\left[ e_{\mu}e^{\dagger}_{\nu} \right] = 2^{2n-1}\delta_{\mu\nu}$.
The relation of convenient for calculations is
\begin{equation}
e_{\mu}e^{\dagger}_{\nu}+e_{\nu}e^{\dagger}_{\mu} = e^{\dagger}_{\mu}e_{\nu}+e^{\dagger}_{\nu}e_{\mu} = 2\delta_{\mu\nu}\mathbf{1}_{2^{2n-1}}.	\label{eq:ASD_basis_prop}
\end{equation}
The (anti-)self-dual tensor $\Sigma^{(\pm)}_{\mu\nu}$ satisfies the $4n$-dimensional (anti-)self-dual relation:
\begin{equation}
\Sigma^{(\pm)}_{[\mu_1\mu_2}\dots \Sigma^{(\pm)}_{\mu_{2n-1}\mu_{2n}]} = \pm\frac{1}{(2n)!}\varepsilon_{\mu_1\mu_2\dots\mu_{2n-1}\mu_{2n}\nu_1\nu_2\dots\nu_{2n-1}\nu_{2n}}\Sigma^{(\pm)}_{\nu_1\nu_2}\dots \Sigma^{(\pm)}_{\nu_{2n-1}\nu_{2n}},	\label{eq:4nd-ASD_relation_tensors}
\end{equation}
where the upper script sign of $\Sigma^{(\pm)}_{\mu\nu}$ corresponds to the sign in the r.h.s. of \eqref{eq:4nd-ASD_relation_tensors}.
For later convenience, we calculate the following quantities:
\begin{align}
\Sigma^{(\pm)}_{12}\dots\Sigma^{(\pm)}_{(4n-1)(4n)} &= \pm(-1)^n2^{2n}\mathbf{1}_{2^{2n-1}},	\notag \\
\Sigma^{(\pm)}_{\mu_1\mu_2}\dots\Sigma^{(\pm)}_{\mu_{4n-1}\mu_{4n}} &= \varepsilon_{\mu_1\mu_2\dots\mu_{4n-1}\mu_{4n}}\Sigma^{(\pm)}_{12}\dots\Sigma^{(\pm)}_{(4n-1)(4n)} = \pm(-1)^n2^{2n}\varepsilon_{\mu_1\mu_2\dots\mu_{4n-1}\mu_{4n}}\mathbf{1}_{2^{2n-1}}.	\label{eq:multi-product_ASD_tensor}
\end{align}

The field strength $F_{\mu\nu}$ of the BPST type instantons is evaluated to be
\begin{equation}
F_{\mu\nu} = \frac{\lambda^2}{(\lambda^2+\|\tilde{x}\|^2)^2}\Sigma^{(\pm)}_{\mu\nu}.	\label{eq:field_strength_BPST}
\end{equation}
Then the field strength \eqref{eq:field_strength_BPST} manifestly satisfies the $4n$-dimensional (anti-)self-dual equation \eqref{eq:ASD_eq_components} by using \eqref{eq:4nd-ASD_relation_tensors}.
The (anti-)self-dual tensor $\Sigma^{(\pm)}_{\mu\nu}$ satisfies the commutation relation:
\begin{equation}
\left[ \Sigma^{(\pm)}_{\mu\nu},\Sigma^{(\pm)}_{\rho\sigma} \right] = 4\left( \delta_{\nu\rho}\Sigma^{(\pm)}_{\mu\sigma} - \delta_{\nu\sigma}\Sigma^{(\pm)}_{\mu\rho} + \delta_{\mu\rho}\Sigma^{(\pm)}_{\sigma\nu} - \delta_{\mu\sigma}\Sigma^{(\pm)}_{\rho\nu} \right).
\end{equation}
Hence, we find that $\Sigma^{(\pm)}_{\mu\nu}$ is the spinor-representation of the SO$(4n)$ Lie algebra.
Therefore the gauge group of the $4n$-dimensional BPST type instanton is the special orthogonal group SO$(4n)$ and its homotopy group is $\pi_{4n-1}(\text{SO}(4n))=\mathbb{Z}\oplus\mathbb{Z}$.
Note that it is sufficient that the homotopy group contains at least one $\mathbb{Z}$ factor to classify instantons by the integer topological charge.

Next, we determine the normalization constant $\mathcal{N}_n$. This is defined by the condition that the topological charge of the BPST instanton \eqref{eq:field_strength_BPST} becomes one.
The topological charge $Q$ of the $4n$-dimensional instantons is defined by the $2n$-th Chern number:
\begin{equation}
Q = (-1)^n\mathcal{N}_n\int_{\mathbb{R}^{4n}}\text{Tr}F(2n) = (-1)^n\mathcal{N}_n\int_{\mathbb{R}^{4n}}d^{4n}x\text{Tr}\left[ \left(\frac{1}{2}\right)^{2n}\varepsilon_{\mu_1\mu_2\dots\mu_{4n-1}\mu_{4n}}F_{\mu_1\mu_2}\dots F_{\mu_{4n-1}\mu_{4n}} \right].
\end{equation}
We easily calculate the topological charge of the BPST type instantons \eqref{eq:field_strength_BPST} by using \eqref{eq:multi-product_ASD_tensor}.
The result is
\begin{align}
Q &= (-1)^n\mathcal{N}_n\frac{1}{2^{2n}}\int_{\mathbb{R}^{4n}}d^{4n}x\left( \frac{\lambda^2}{(\lambda^2+\|\tilde{x}\|^2)^2} \right)^{2n}\text{Tr}\left[ \varepsilon_{\mu_1\mu_2\dots\mu_{4n-1}\mu_{4n}}\Sigma^{(\pm)}_{\mu_1\mu_2}\dots\Sigma^{(\pm)}_{\mu_{4n-1}\mu_{4n}} \right]	\notag \\
&=\pm\frac{2^{2n}(4n)!n\pi^{2n}}{\Gamma(2n+1)}B(2n,2n)\mathcal{N}_n = \pm(2n)!(2\pi)^{2n}\mathcal{N}_n,
\end{align}
where $B(2n,2n)$ is the beta function and we have used the following relation:
\begin{equation}
\int_0^{\infty}dr\frac{r^{4n-1}}{(1+r^2)^{4n}} = \frac{1}{2}B(2n,2n).
\end{equation}
We define that the topological charge of the instantons is a positive number when the instantons satisfy the self-dual equation, i.e. the plus sign in \eqref{eq:ASD_eq_components}.
Therefore the $4n$-dimensional normalization constants $\mathcal{N}_n$ is determined to be 
\begin{equation}
\mathcal{N}_n = \frac{1}{(2n)!(2\pi)^{2n}}.
\end{equation}

\section{U$(N)$ ADHM construction in $4n$ dimensions ($n\geq2$)}
In this section, we study an ADHM construction of the (anti-)self-dual instanton in the $4n$-dimensional Euclidean space with the flat metric.
In the following, we choose the anti-self-dual solutions to the equation \eqref{eq:ASD_eq_components} and we use the matrix representation of the Clifford algebra $C\ell_{4n-1}(\mathbb{C})$. This explicit form can be found in Appendix A. We first introduce the $4n$-dimensional Weyl operator:
\begin{equation}
\Delta = C(x\otimes \mathbf{1}_k) + D,	\label{eq:Weyl_op}
\end{equation}
where $x=x^{\mu}e_{\mu}$, the symbol $\otimes$ means the tensor product, $C$ and $D$ are $(N+2^{2n-1}k)\times 2^{2n-1}k$ constant matrices which are called the ADHM data, and $N$ corresponds to the rank of the unitary group, we will show this fact for later.
If we consider self-dual solutions then we must choose the basis $e^{\dagger}_{\mu}$ instead of $e_{\mu}$.
In the next section, we will show that the integer $k$ corresponds to the instanton number which is defined by the $2n$-th Chern number $k=|\mathcal{N}_n\int\text{Tr}F(2n)|$.
Now we demand that the Weyl operator satisfies the first ADHM constraint:
\begin{equation}
\Delta^{\dagger}\Delta = \mathbf{1}_{2^{2n-1}}\otimes E_k^{(1)},	\label{eq:first_ADHM_constraint}
\end{equation}
and the second ADHM constraint:
\begin{equation}
C^{\dagger}\Delta\left( \Delta^{\dagger}\Delta \right)^{-1}\Delta^{\dagger}C = \mathbf{1}_{2^{2n-1}}\otimes E_k^{(2)},	\label{eq:second_ADHM_constraint}
\end{equation}
where $\Delta^{\dagger}$ is the Hermitian conjugate matrix of $\Delta$, $E_k^{(1)}$ and $E_k^{(2)}$ are invertible $k\times k$ matrices.
The first ADHM constraint \eqref{eq:first_ADHM_constraint} is the natural generalization of the four-dimensional one \cite{Atiyah:1978ri}.
On the other hand, the second ADHM constraint \eqref{eq:second_ADHM_constraint} is the analogy of the eight-dimensional one \cite{Nakamula:2016srw}.
In addition, the Weyl operator requires the non-degeneracy condition: $\text{rank}~\Delta=2^{2n-1}k$,
and the existence of the inverse $E^{(a)}_k~(a=1,2)$ is guaranteed by this condition (This proof is shown in Appendix B).
Here rank $A$ means the rank of the matrix $A$,
and the non-degeneracy condition of the Weyl operator is satisfied if and only if the ADHM data $C,D$ satisfy the condition: $\text{rank}~C=\text{rank}~D=2^{2n-1}k$.
For later convenience, let us analyze the ADHM constraints in more detail. 
For \eqref{eq:Weyl_op}, the first ADHM constraint \eqref{eq:first_ADHM_constraint} becomes
\begin{equation}
(x^{\dagger}\otimes\mathbf{1}_k)C^{\dagger}C(x\otimes\mathbf{1}_k) + (x^{\dagger}\otimes\mathbf{1}_k)C^{\dagger}D + D^{\dagger}C(x\otimes\mathbf{1}_k) + D^{\dagger}D = \mathbf{1}_{2^{2n-1}}\otimes E_k^{(1)}(x).
\end{equation}
The ADHM constraints hold for all $x\in\mathbb{R}^{4n}$, hence we can decompose the first ADHM constraint to three $x$-independent conditions:
\begin{subequations}
\begin{align}
C^{\dagger}C &= \mathbf{1}_{2^{2n-1}}\otimes E_k^{(1,1)},	\label{eq:first_ADHM_x-independence_1} \\
C^{\dagger}D &= e_{\mu}\otimes E_{k,\mu}^{(1,2)},	\label{eq:first_ADHM_x-independence_2} \\
D^{\dagger}D &= \mathbf{1}_{2^{2n-1}}\otimes E_k^{(1,3)},	\label{eq:first_ADHM_x-independence_3}
\end{align}	\label{eq:first_ADHM_x-independence}
\end{subequations}
where $E_{k,\mu}^{(1,2)}$ is a hermite matrix and $E_k^{(1)} = x^2E_k^{(1,1)} + 2x^{\mu}E_{k,\mu}^{(1,2)} + E_k^{(1,3)}.$
Similarly, the second ADHM constraint \eqref{eq:second_ADHM_constraint} expands to
\begin{align}
&C^{\dagger}C(x\otimes\mathbf{1}_k)(\mathbf{1}_{2^{2n-1}}\otimes f)(x^{\dagger}\otimes\mathbf{1}_k)C^{\dagger}C + C^{\dagger}C(x\otimes\mathbf{1}_k)(\mathbf{1}_{2^{2n-1}}\otimes f)D^{\dagger}C	\notag \\
&\hspace{60pt}+C^{\dagger}D(\mathbf{1}_{2n-1}\otimes f)(x^{\dagger}\otimes\mathbf{1}_k)C^{\dagger}C + C^{\dagger}D(\mathbf{1}_{2^{2n-1}}\otimes f)D^{\dagger}C = \mathbf{1}_{2^{2n-1}}\otimes E_k^{(2)}(x),	\label{eq:second_ADHM_expanded}
\end{align}
where $f^{-1}=E_k^{(1)}$. 
When we discuss the $x$-independent conditions of the second ADHM constraint, we can ignore the $x$ that is included in $f$ because the matrix $f$ is already placed in the r.h.s. for the tensor product.
For \eqref{eq:first_ADHM_x-independence_1} and $x^{\dagger}x=xx^{\dagger}=x^2\mathbf{1}_{2^{2n-1}}$, the $x^2$ term in \eqref{eq:second_ADHM_expanded} automatically satisfies the constraint.
The $x^1$ terms in \eqref{eq:second_ADHM_expanded} expands to
\begin{align}
&C^{\dagger}C(x\otimes\mathbf{1}_k)(\mathbf{1}_{2^{2n-1}}\otimes f)D^{\dagger}C +C^{\dagger}D(\mathbf{1}_{2n-1}\otimes f)(x^{\dagger}\otimes\mathbf{1}_k)C^{\dagger}C \notag \\
&\hspace{100pt} = x^{\nu}\left( e_{\nu}e_{\mu}^{\dagger}\otimes E_k^{(1,1)}fE_{k,\mu}^{(1,2)} + e_{\mu}e_{\nu}^{\dagger}\otimes E_{k,\mu}^{(1,2)}fE_k^{(1,1)} \right).
\end{align}
These terms satisfy the constraint for all $x$ if and only if the following condition holds:
\begin{equation}
E_k^{(1,1)}fE_{k,\mu}^{(1,2)} = E_{k,\mu}^{(1,2)}fE_k^{(1,1)}.	\label{eq:second_ADHM_x-independence_1}
\end{equation}
Next we consider the $x^0$ term in \eqref{eq:second_ADHM_expanded}.
This term becomes
\begin{equation}
C^{\dagger}D(\mathbf{1}_{2^{2n-1}}\otimes f)D^{\dagger}C = \left( \delta_{\mu\nu}\mathbf{1}_{2^{2n-1}}+\Sigma^{(-)}_{\mu\nu}/2 \right)\otimes E_{k,\mu}^{(1,2)}fE_{k,\nu}^{(1,2)},
\end{equation}
here we have used $e_{\mu}e_{\nu}^{\dagger}=\delta_{\mu\nu}\mathbf{1}_{2^{2n-1}}+\Sigma^{(-)}_{\mu\nu}/2$.
For this equation, we obtain the following condition for the $x^0$ term:
\begin{equation}
\Sigma^{(-)}_{\mu\nu} \otimes E_{k,\mu}^{(1,2)}fE_{k,\nu}^{(1,2)} = 0.	\label{eq:second_ADHM_x-independence_2}
\end{equation}
Therefore we obtained the two $x$-independent conditions of the second ADHM constraint, namely \eqref{eq:second_ADHM_x-independence_1} and \eqref{eq:second_ADHM_x-independence_2}.

Let us show that how to obtain the gauge field of the anti-self-dual instanton from the ADHM data.
Following the ADHM construction in four dimensions \cite{Atiyah:1978ri}, we first consider zero modes of the Weyl operator $\Delta$.
The null-space of the Hermitian conjugate matrix $\Delta^{\dagger}$ is $N$-dimensional, as it has $N$ fewer rows than columns.
The basis vectors for this null-space can be assembled into an $(N+2^{2n-1}k)\times N$ matrix $V(x)$, which is sometimes called the zero mode.
This fact means that the zero mode $V(x)$ is the solution to the Weyl equation:
\begin{equation}
\Delta^{\dagger}V(x)=0,	\label{eq:Weyl_eq}
\end{equation}
and the zero mode $V(x)$ is normalized as $V^{\dagger}V=\mathbf{1}_N$.
The zero mode $V$ and the Weyl operator $\Delta$ satisfy the following relation which is called the completeness relation:
\begin{equation}
\mathbf{1}_{N+2^{2n-1}k} - VV^{\dagger} = \Delta(\Delta^{\dagger}\Delta)^{-1}\Delta^{\dagger}.	\label{eq:completeness_relation}
\end{equation}
We can easily prove this relation by using a $(N+2^{2n-1}k)\times(N+2^{2n-1}k)$ matrix $W=\begin{pmatrix} \Delta & V \end{pmatrix}$.
Because of the non-degeneracy condition, the Weyl equation \eqref{eq:Weyl_eq} and the normalization: $V^{\dagger}V=\mathbf{1}_N$, the columns of $W$ are linearly independent.
Therefore the matrix $W$ is invertible, and the following equation is an identity equation: $W(W^{\dagger}W)^{-1}W^{\dagger} = \mathbf{1}_{N+2^{2n-1}k}$.
We can obtain the completeness relation by expanding the l.h.s. term $W(W^{\dagger}W)^{-1}W^{\dagger}$ with $\Delta$ and $V$.
We employ the ansatz of the gauge field $A_{\mu}(x)$ is given by the pure gauge form:
\begin{equation}
A_{\mu}(x) = V^{\dagger}(x)\partial_{\mu}V(x).	\label{eq:ADHM_pure_gauge}
\end{equation}

Next we confirm that the field strength $F_{\mu\nu}$ from the ansatz \eqref{eq:ADHM_pure_gauge} automatically satisfies the anti-self-dual equation \eqref{eq:ASD_eq_components}.
For the Weyl equation \eqref{eq:Weyl_eq} and the completeness relation \eqref{eq:completeness_relation}, the field strength becomes
\begin{equation}
F_{\mu\nu} = V^{\dagger}C\left( e_{\mu}\otimes\mathbf{1}_k \right)\left( \Delta^{\dagger}\Delta \right)^{-1}\left( e^{\dagger}_{\nu}\otimes\mathbf{1}_k \right)C^{\dagger}V - (\mu\leftrightarrow\nu).
\end{equation}
Now we use the first ADHM constraint \eqref{eq:first_ADHM_constraint} then the factor $(\Delta^{\dagger}\Delta)^{-1}$ commutes with the basis $e_{\mu}\otimes\mathbf{1}_k$.
Hence the field strength becomes
\begin{equation}
F_{\mu\nu} = V^{\dagger}C(\Delta^{\dagger}\Delta)^{-1}\left( \Sigma^{(-)}_{\mu\nu}\otimes\mathbf{1}_k \right)C^{\dagger}V.
\end{equation}
Therefore the multi-product of the field strengths is
\begin{equation}
F_{\mu_1\mu_2}\dots F_{\mu_{2n-1}\mu_{2n}} = \left( V^{\dagger}C(\Delta^{\dagger}\Delta)^{-1}\left( \Sigma^{(-)}_{\mu_1\mu_2}\otimes\mathbf{1}_k \right)C^{\dagger}V \right)\dots\left( V^{\dagger}C(\Delta^{\dagger}\Delta)^{-1}\left( \Sigma^{(-)}_{\mu_{2n-1}\mu_{2n}}\otimes\mathbf{1}_k \right)C^{\dagger}V \right).	\label{eq:ADHM_F^n}
\end{equation}
We order that $\Sigma^{(-)}_{\mu\nu}\otimes\mathbf{1}_k$ commute with $C^{\dagger}VV^{\dagger}C$ in \eqref{eq:ADHM_F^n},
thus we demand the following condition:
\begin{equation}
e_{\mu}\otimes\mathbf{1}_k\left( C^{\dagger}VV^{\dagger}C \right) = \left( C^{\dagger}VV^{\dagger}C \right)e_{\mu}\otimes\mathbf{1}_k.	\label{eq:pre_ADHM_constraint}
\end{equation}
Now we use the completeness relation \eqref{eq:completeness_relation} then the condition \eqref{eq:pre_ADHM_constraint} is decomposed as
\begin{equation}
e_{\mu}\otimes\mathbf{1}_k\left( C^{\dagger}C \right) = \left( C^{\dagger}C \right)e_{\mu}\otimes\mathbf{1}_k,~~
e_{\mu}\otimes\mathbf{1}_k\left( C^{\dagger}\Delta(\Delta^{\dagger}\Delta)^{-1}\Delta^{\dagger}C \right) = \left( C^{\dagger}\Delta(\Delta^{\dagger}\Delta)^{-1}\Delta^{\dagger}C \right)e_{\mu}\otimes\mathbf{1}_k.	\label{eq:pre-second_ADHM_decomposed}
\end{equation}
For \eqref{eq:first_ADHM_x-independence_1}, the first condition is automatically satisfied when the first ADHM constraint \eqref{eq:first_ADHM_constraint} holds.
On the other hand, the second condition is just the second ADHM constraint \eqref{eq:second_ADHM_constraint}.
We find that the condition \eqref{eq:pre_ADHM_constraint} is satisfied when the first and second ADHM constraints hold.
Therefore, for the condition \eqref{eq:pre_ADHM_constraint}, the multi-product of the field strengths becomes
\begin{equation}
F_{\mu_1\mu_2}\dots F_{\mu_{2n-1}\mu_{2n}} = V^{\dagger}C\left( \Delta^{\dagger}\Delta \right)^{-1}\left( \Sigma^{(-)}_{\mu_1\mu_2}\dots\Sigma^{(-)}_{\mu_{2n-1}\mu_{2n}} \otimes \mathbf{1}_k \right)C^{\dagger}V\left( V^{\dagger}C\left( \Delta^{\dagger}\Delta \right)^{-1}C^{\dagger}V \right)^{n-1}.	\label{eq:Fn_components_ADHM_form}
\end{equation}
Since $\Sigma^{(-)}_{\mu_1\mu_2}\dots\Sigma^{(-)}_{\mu_{2n-1}\mu_{2n}}$ satisfies the anti-self-dual relation \eqref{eq:4nd-ASD_relation_tensors}, we have shown that the field strengths $F_{\mu\nu}$ that are constructed from the $4n$-dimensional ADHM construction satisfy the anti-self-dual equation in $4n$ dimensions \eqref{eq:ASD_eq_components}.

We show that the ADHM data can transform more simplify form without loss of generality.
It is easy to find that the Weyl equation \eqref{eq:Weyl_eq}, the normalization condition $V^{\dagger}V=\mathbf{1}_N$, the first and second ADHM constraints \eqref{eq:first_ADHM_constraint},\eqref{eq:second_ADHM_constraint} are invariant under the following transformations:
\begin{equation}
C\mapsto C'=\mathcal{U}C\mathcal{R},~~D\mapsto D'=\mathcal{U}D\mathcal{R},~~V\mapsto V'=\mathcal{U}V,	\label{eq:ADHM_data_transformation}
\end{equation}
where $\mathcal{U}\in\text{U}(N+2^{2n-1}k)$ and $\mathcal{R} = \mathbf{1}_{2^{2n-1}}\otimes\mathcal{R}_k \in \mathbf{1}_{2^{2n-1}}\otimes \text{GL}(k;\mathbb{C})$.
Using this $\text{U}(N+2^{2n-1}k)\times \text{GL}(k;\mathbb{C})$ transformation, we can fix the ADHM data to the so-called a canonical form:
\begin{equation}
C=
\begin{pmatrix}
0_{[N]\times[2^{2n-1}k]} \\ \mathbf{1}_{2^{2n-1}k}
\end{pmatrix},~~~~
D=
\begin{pmatrix}
S_{[N]\times[2^{2n-1}k]}	\\ T_{[2^{2n-1}k]\times[2^{2n-1}k]}
\end{pmatrix}.	\label{eq:ADHM_data_canonical_form}
\end{equation}
Here the matrix subscript $[a]\times[b]$ means the matrix size, and the symbol $S_{[N]\times[2^{2n-1}k]}$ stands for 
$\begin{pmatrix} 
S_{1~[N]\times[k]} & \dots & S_{2^{2n-1}~[N]\times[k]}
 \end{pmatrix}$.
The existence of the canonical form is guaranteed by the non-degeneracy condition.
In the canonical form, all the ADHM data are included in the matrices $S$ and $T$.
Let us now rewrite the $x$-independent conditions of the first and second ADHM constraints  \eqref{eq:first_ADHM_x-independence},\eqref{eq:second_ADHM_x-independence_1},\eqref{eq:second_ADHM_x-independence_2} in the canonical form.
In this case, $C^{\dagger}C=\mathbf{1}_{2^{2n-1}k}=\mathbf{1}_{2^{2n-1}}\otimes\mathbf{1}_k$ thus the condition \eqref{eq:first_ADHM_x-independence_1} is automatically satisfied.
The condition \eqref{eq:first_ADHM_x-independence_2} means that the matrix $C^{\dagger}D$ is written with the (anti-)self-dual basis $e_{\mu}$.
In the canonical form, $C^{\dagger}D=T$ thus \eqref{eq:first_ADHM_x-independence_2} becomes
\begin{equation}
T=e_{\mu}\otimes T^{\mu},
\end{equation}
where $T^{\mu}$ is a $k\times k$ hermite matrix.
The condition \eqref{eq:first_ADHM_x-independence_3} is rewritten as
\begin{equation}
S^{\dagger}S + T^{\dagger}T = \mathbf{1}_{2^{2n-1}}\otimes E_k^{(1,3)}.	\label{eq:first_ADHM_x-independence_CF}
\end{equation}
This $x$-independent condition is the natural generalization of four-dimensional one which is usually called the ADHM equation.
On the other hand, the $x$-independent conditions of the second ADHM constraint lead to new type ADHM equations.
In the canonical form, $E_k^{(1,1)}=\mathbf{1}_k$ and $\delta^{\mu\nu}E_{k,\nu}^{(1,2)}=T^{\mu}$ thus the condition \eqref{eq:second_ADHM_x-independence_1} becomes
\begin{equation}
fT^{\mu}=T^{\mu}f.	\label{eq:first_ADHM_x-independence_CF_1}
\end{equation}
For this condition, the condition \eqref{eq:second_ADHM_x-independence_2} is rewritten as
\begin{equation}
\Sigma^{(-)}_{\mu\nu}\otimes T^{\mu}T^{\nu} =0.	\label{eq:first_ADHM_x-independence_CF_2}
\end{equation}
In higher dimensions, the ADHM data must satisfy the new type ADHM equations \eqref{eq:first_ADHM_x-independence_CF_1} and \eqref{eq:first_ADHM_x-independence_CF_2} in addition to the standard type one \eqref{eq:first_ADHM_x-independence_CF}.
Finally, we note that there are residual symmetries which leave the canonical form \eqref{eq:ADHM_data_canonical_form} invariant.
The transformations are given by
\begin{equation}
S_a\mapsto S'_a=QS_aR,~~ T^{\mu}\mapsto T'^{\mu}=R^{\dagger}T^{\mu}R,
\end{equation}
where the index $a$ runs from $1$ to $2^{2n-1}$, $Q\in \text{SU}(N)$ and $R\in\text{U}(k)$.

Next, we study the gauge group of the instantons that are generated from the ADHM construction.
The transformation of the zero mode which preserves the normalization condition $V^{\dagger}V=\mathbf{1}_N$ is given by
\begin{equation}
V(x)\mapsto V(x) g(x),~~~g(x)\in \text{U}(N).
\end{equation}
Note that this transformation is independent of the transformation \eqref{eq:ADHM_data_transformation}.
This zero mode transformation leads to a gauge field transformation through \eqref{eq:ADHM_pure_gauge}.
Indeed, the gauge field is transformed to
\begin{equation}
A_{\mu} \mapsto g^{-1}(x)A_{\mu}g(x) + g^{-1}(x)\partial_{\mu}g(x).
\end{equation}
This transformation is same as the ordinary gauge transformation.
Hence, the instantons that are generated from the ADHM construction possess the unitary gauge group U$(N)$.
Because of this fact, in the special case $k=0$, we find that the ansatz \eqref{eq:ADHM_pure_gauge} gives a pure gauge, namely, it automatically solves the (anti-)self-dual equation \eqref{eq:ASD_eq_components} in the vacuum sector.
We are interested in instantons that are characterized by the instanton number $k$, but the homotopy groups become trivial when the rank of the unitary group is small.
Therefore the rank of the gauge group $N$ is restricted by the condition that the homotopy group $\pi_{4n-1}(\text{U}(N))$ is non-trivial.
The non-trivial homotopy groups of the (special) unitary group are
\begin{equation}
\pi_{4n-1}(\text{U}(N)) = \pi_{4n-1}(\text{SU}(N)) = \mathbb{Z},~~~~N\geq 2n.
\end{equation}
From this fact, we demand the condition $N\geq2n$ when we consider the topological instantons.
In addition, we note that the ADHM construction does not impose the speciality condition on the gauge group in general, namely the gauge group is not the special unitary group SU$(N)$.
We can decompose the group U$(N)$ into the special group SU$(N)$ part and U$(1)$ part: $\text{U}(N)=\text{SU}(N)\ltimes\text{U}(1)$.
Here the symbol $\ltimes$ means the semidirect product of the group.
Usually, we must fix the element of U$(1)$ by hand when we consider SU$(N)$ instantons in the ADHM construction.

For later convenience, we show a formula of the topological charge density.
The topological charge $Q$ for the $4n$-dimensional instantons is defined by the $2n$-th Chern number $Q = \mathcal{N}_n\int\text{Tr}F(2n)$.
Now we define the charge density $\mathcal{Q}$ as $Q=\mathcal{N}_n\int d^{4n}x\mathcal{Q}$,
then using the expression \eqref{eq:Fn_components_ADHM_form}, the ADHM constraints \eqref{eq:first_ADHM_constraint}, \eqref{eq:second_ADHM_constraint}, and the multi-product of the (anti-)self-dual tensors \eqref{eq:multi-product_ASD_tensor},
we obtain the charge density formula:
\begin{equation}
\mathcal{Q} = \pm(-1)^n(4n)!\text{Tr}\left( V^{\dagger}C(\Delta^{\dagger}\Delta)^{-1}C^{\dagger}V \right)^{2n}.	\label{eq:charge_density_formula}
\end{equation}
Here $\pm$ corresponds to the (anti-)self-dual solution (tensor) respectively.
When the ADHM data is the canonical form \eqref{eq:ADHM_data_canonical_form}, we can rewrite \eqref{eq:charge_density_formula} as
\begin{equation}
\mathcal{Q} = \pm(-1)^n2^{2n-1}(4n)!\text{Tr}_k\left( \left( E^{(1)}_k \right)^{-1}\left( \mathbf{1}_k - E^{(2)}_k \right) \right)^{2n}.	\label{eq:charge_density_formula_2}
\end{equation}

Now we have introduced the ADHM construction of the (anti-)self-dual instantons in $4n$ dimensions.
Here we have some comments on the higher dimensional ADHM construction.
Compared with the four-dimensional ADHM construction, the first ADHM constraints \eqref{eq:first_ADHM_constraint} is the natural generalization of the four-dimensional one.
On the other hand, the second ADHM constraint \eqref{eq:second_ADHM_constraint} is an essentially new constraint and this new constraint corresponds to the non-linearity of the (anti-)self-dual equation \eqref{eq:ASD_eq_components}.
It is difficult to construct the multi-instantons in higher dimensions, because of the non-linearity of the (anti-)self-dual equations.
In the higher dimensional ADHM construction is similar to this situation, namely the constructions of the multi-instantons are difficult by the second ADHM constraint.
We will discuss this fact in more detail in the next section.

\section{Higher-dimensional ADHM data with U$(2^{2n-1})$ gauge group}
In this section, we introduce explicit ADHM data in higher dimensions ($n\geq2$).
However, it is hard that we find an essentially new ADHM data, hence we will consider the data type that is generalizing the well known four-dimensional one and choose the rank of the gauge group to $N=2^{2n-1}$.
Here we recall that the first ADHM constraint is the natural generalization of the four-dimensional one. 
Therefore the data type that is generalizing the four-dimensional ADHM data already satisfies the first ADHM constraint, and we call this data type as an ADHM ``ansatz''.

The second ADHM constraint \eqref{eq:second_ADHM_constraint} contains the inverse matrix $(\Delta^{\dagger}\Delta)^{-1}$, hence the calculation of this constraint is hard in general.
Therefore we use the following constraint instead of the second ADHM constraint to confirm that the ADHM ansatz is well-defined as a higher-dimensional ADHM data:
\begin{equation}
C^{\dagger}VV^{\dagger}C = \mathbf{1}_{2^{2n-1}}\otimes E^{(3)}_k,	\label{eq:pre_second_ADHM_constraint}
\end{equation}
where $E^{(3)}_k$ is an invertible $k\times k$ matrix.
The existence of the inverse $E^{(3)}_k$ is guaranteed by that $E^{(1)}_k$ and $E^{(2)}_k$ are invertible.
Although this constraint contains the zero mode $V$, the calculation of the Weyl equation \eqref{eq:Weyl_eq} is more easily than the calculation of the inverse matrix $(\Delta^{\dagger}\Delta)^{-1}$ in general, namely we can calculate the constraint \eqref{eq:pre_second_ADHM_constraint} more easily than the second ADHM constraint.
This constraint is same as the condition \eqref{eq:pre_ADHM_constraint}, therefore the ADHM ansatz satisfy the first ADHM constraint then this constraint is same as the second ADHM constraint for \eqref{eq:pre-second_ADHM_decomposed}.

\subsection{BPST type one-instanton}
In the case of $k=1$, the ADHM ansatz in the canonical form is the simplest one:
\begin{equation}
C=
\begin{pmatrix}
0 \\ \mathbf{1}_{2^{2n-1}}
\end{pmatrix},~~
D=
\begin{pmatrix}
\lambda\mathbf{1}_{2^{2n-1}} \\ -a^{\mu}e_{\mu}
\end{pmatrix},	\label{eq:BPST_ADHM_data}
\end{equation}
where $\lambda\in\mathbb{R}$ is the size modulus and $a^{\mu}\in\mathbb{R}$ is the position modulus of the instanton.
The solution to the Weyl equation \eqref{eq:Weyl_eq} is
\begin{equation}
V = \frac{1}{\sqrt{\rho}}
\begin{pmatrix}
\tilde{x}^{\dagger} \\ -\lambda\mathbf{1}_{2^{2n-1}}
\end{pmatrix},
\end{equation}
where $\tilde{x}^{\dagger}=(x^{\mu}-a^{\mu})e^{\dagger}_{\mu}$ and $\rho=\lambda^2+\|\tilde{x}\|^2$.
The l.h.s. of the constraint \eqref{eq:pre_second_ADHM_constraint} that is associated with the BPST type ADHM ansatz \eqref{eq:BPST_ADHM_data} is proportional to the identity $\mathbf{1}_{2^{2n-1}}$:
\begin{equation}
C^{\dagger}VV^{\dagger}C = \frac{\lambda^2}{\rho}\mathbf{1}_{2^{2n-1}}.
\end{equation}
Hence, this ansatz \eqref{eq:BPST_ADHM_data} is well-defined as the ADHM data of the anti-self-dual one-instanton.
Indeed, we easily confirm that this ADHM data reproduces the BPST type gauge field \eqref{eq:BPST_gauge_field} by using \eqref{eq:ADHM_pure_gauge}.

\subsection{'t Hooft type ansatz}
We next consider the ADHM data with higher charges.
A natural candidate for multi-instanton ADHM data is generalization of the 't Hooft type one.
However, in the case of $n=2$ (i.e. eight dimensions), it was shown that the simple generalization of the 't Hooft type ADHM ansatz is not well-defined as the higher-dimensional ADHM data in \cite{Nakamula:2016srw}.
In the following, we will show that the cases in more higher dimensions are same situations as the eight-dimensional case.
The 't Hooft type ADHM ansatz is given by
\begin{align}
T^{\mu} &= \text{diag}_{p=1}^{k}\left( -a^{\mu}_p \right),& S &= \mathbf{1}_{2^{2n-1}}\otimes\begin{pmatrix} \lambda_1 & \lambda_2 & \dots & \lambda_k \end{pmatrix},	\label{eq:tHooft_ADHM_data}
\end{align}
with $a^{\mu}_p\in\mathbb{R}$ is the instanton position and $\lambda_p\in\mathbb{R}$ is the instanton size moduli respectively.
The Weyl operator that is associated with the 't Hooft type ADHM ansatz is
\begin{equation}
\Delta^{\dagger} =
\begin{pmatrix}
S^{\dagger}	& e^{\dagger}_{\mu}\otimes\left( x^{\mu}\mathbf{1}_k + T^{\mu} \right)
\end{pmatrix},
\end{equation}
therefore the solution to the Weyl equation \eqref{eq:Weyl_eq} is
\begin{equation}
V=\frac{1}{\sqrt{\phi}}
\begin{pmatrix}
-\mathbf{1}_{2^{2n-1}}	\\
\left( e_{\mu}\otimes\text{diag}_{p=1}^k\left( \frac{\tilde{x}^{\mu}_p}{\|\tilde{x}_p\|^2} \right)S^{\dagger} \right)
\end{pmatrix}.
\end{equation}
Here we have defined $\phi=1+\sum_{p=1}^k\frac{\lambda_p^2}{\|\tilde{x}_p\|^2}$, $\tilde{x}^{\mu}_p=x^{\mu}-a^{\mu}_p$ and $\|\tilde{x}_p\|^2=\tilde{x}^{\mu}_p\tilde{x}^{\mu}_{p}$ ($p$ is not summed).
For the constraint \eqref{eq:pre_second_ADHM_constraint} that is associated with 't Hooft type ADHM ansatz, we obtain the following condition with the moduli $\lambda_m$ and $a^{\mu}_m$:
\begin{equation}
\lambda_m\lambda_n\left( x^{\mu}-a^{\mu}_m \right)\left( x^{\nu}-a^{\nu}_n \right)\Sigma^{(-)}_{\mu\nu} = 0,	\label{eq:ADHM_data_condition_tHooft}
\end{equation}
where the indices $m,n$ run from $1$ to $k$ and are not summed.
This condition is trivially satisfied in the case of $k=1$, however, for arbitrary moduli parameters $\lambda_m,a_m$, this is not satisfied in the higher charges $k\geq2$.
Therefore the simple generalization of the 't Hooft type ADHM ansatz is not well-defined as the ADHM data with higher charges ($k\geq2$).

Let us now demand the following condition for moduli parameters to satisfy the condition \eqref{eq:ADHM_data_condition_tHooft}:
\begin{equation}
\|a^{\mu}_m-a^{\mu}_n\|^2 \gg \lambda_m\lambda_n,	\label{eq:well-separated_limit}
\end{equation}
for all $m$ and $n$.
This condition means that each instanton is well-separated, hence we call this condition as the well-separated limit or the dilute instanton gas limit (approximation) \cite{Christ:1978jy}.
In the well-separated limit \eqref{eq:well-separated_limit}, we neglect all the off-diagonal components of the matrix $E_k^{(1)}$ in \eqref{eq:first_ADHM_constraint}:
\begin{equation}
E_k^{(1)} =
\begin{pmatrix}
\lambda_1^2+\|\tilde{x}_1\|^2 & \dots & \lambda_1\lambda_k \\
\vdots & \ddots & \vdots \\
\lambda_k\lambda_1 & \dots & \lambda_k^2+\|\tilde{x}_k\|^2
\end{pmatrix}
\simeq
\begin{pmatrix}
 \lambda_1^2+\|\tilde{x}_1\|^2 & \dots & 0 \\
 \vdots & \ddots & \vdots \\
 0 & \dots & \lambda_k^2+\|\tilde{x}_k\|^2
\end{pmatrix}.
\label{eq:R8-tHooft_data_well-separated}
\end{equation}
Thus the second ADHM constraint becomes
\begin{align}
C^{\dagger}\Delta(\Delta^{\dagger}\Delta)^{-1}\Delta^{\dagger} C
\simeq \mathbf{1}_{2^{2n-1}}\otimes \text{diag}_{p=1}^k\left[
 \frac{\|\tilde{x}_p\|^2}{\lambda_p^2+\|\tilde{x}_p\|^2} \right].
\label{eq:8-dim_well-separated_approximate_additional_ADHM_constraint}
\end{align}
Therefore the 't Hooft type ADHM ansatz \eqref{eq:tHooft_ADHM_data} in the well-separated limit satisfies the second ADHM constraint \eqref{eq:second_ADHM_constraint}.

Some comments are in order.
First, we can find exact solutions to the condition \eqref{eq:ADHM_data_condition_tHooft}, but these solutions are unsuitable data for multi-instantons.
The condition is exactly solved by $\lambda_m=0~(\text{for all $m$})$, but this solution makes the pure gauge field, namely, it is a vacuum configuration.
On the other hand, we find another exact solution $a_m=a_n~(m\neq n)$ which means that all the instantons are localized at the same point. However, this solution is equivalent to the one-instanton's one.

Second, we will show that the topological charge of the 't Hooft type $k$-instantons in the well-separated limit is an integer.
For $k=1$, the charge density is calculated with using \eqref{eq:charge_density_formula_2}:
\begin{equation}
\mathcal{Q}^{(k=1)}_{\text{'t Hooft}} = -(-1)^n2^{2n-1}(4n)!\left( \frac{\lambda^2}{\left( \lambda^2+\|\tilde{x}\|^2 \right)^2} \right)^{2n}.
\end{equation}
This is same as the BPST one anti-instanton charge density, hence we find that $|Q^{(k=1)}_{\text{'t Hooft}}|=1$.
In the case of $k\geq 2$, for the charge density formula \eqref{eq:charge_density_formula_2}, the charge density of the 't Hooft type in the well-separated limit is given by
\begin{equation}
\mathcal{Q}_{\text{'t Hooft}} \simeq -(-1)^n2^{2n-1}(4n)!\sum_{p=1}^k\left( \frac{\lambda_p^2}{\left( \lambda_p^2+\|\tilde{x}_p\|^2 \right)^2} \right)^{2n}.
\end{equation}
This is the summation of the above mentioned one-instanton charge density, therefore we obtain $|Q_{\text{'t Hooft}}|=k$.

Finally, we mention a higher-dimensional Jackiw-Nohl-Rebbi (JNR) type ADHM data.
The JNR type ADHM data in four dimensions \cite{Corrigan:1978ce} is well known as multi-instantons data which is different from the 't Hooft type. 
We can give generalization of the higher-dimensional JNR type ansatz, for instance, an eight-dimensional case in \cite{Nakamula:2016srw}. However, the second ADHM constraint requires the same condition as 't Hooft one \eqref{eq:ADHM_data_condition_tHooft}.
Therefore the JNR type ADHM data is well-defined only if we assume the well-separated limit.

\section{Calorons in $4n$ dimensions and the monopole limit}
In this section, we consider higher-dimensional calorons and the monopole limit.
It is well known that the Harrington-Shepard (HS) one-caloron in the four dimensions can be generated by the 't Hooft multi-instantons that are periodic in one of the four coordinates \cite{Harrington:1978ve}.
Can we generate a higher-dimensional HS type one-caloron with the same method in the four dimensions?
Let us discuss this question in the following.
We will use the multi-instantons to produce the HS type caloron.
However, the 't Hooft type multi-instantons in the higher dimensions are well-defined only if we assume the well-separated limit \eqref{eq:well-separated_limit}.
Therefore we use the 't Hooft type multi-instanton with well-separated on the periodic coordinate direction $t=x^{4n}$ to produce the higher-dimensional caloron.

We consider the situation that same size 't Hooft type one-instantons are lined up on the $x^{4n}$-direction with well-separated.
This gauge field is given by
\begin{equation}
A_{\mu}(x) = \frac{1}{4}\Sigma^{(\pm)}_{\mu\nu}\partial_{\nu}\ln\phi_{\text{'t Hooft}}(x),
\end{equation}
where 
\begin{equation}
\phi_{\text{'t Hooft}}(x) = 1+\sum_{p=-P}^P\frac{\lambda^2}{\|\mathbf{x}-\mathbf{a}_x\|^2+(x^{4n}-a^{4n}_p)^2}.
\end{equation}
Here $\lambda\in\mathbb{R}$ is the instanton size, $\mathbf{x}=x^i~(i=1,2,\dots,4n-1)$, $\mathbf{a}_x\in\mathbb{R}^{4n-1}$ is the instanton's position (without the $x^{4n}$-direction) and $a^{4n}_p\in\mathbb{R}$ is the positions on the $x^{4n}$-direction.
For the well-separated limit \eqref{eq:well-separated_limit}, the $x^{4n}$-direction position $a^{4n}_p$ satisfies the condition: $(a^{4n}_p-a^{4n}_q)^2\gg \lambda^2~(p\neq q)$.

Now we choose the $x^{4n}$-direction positions $a^{4n}_p=a_t - p\beta~(a_t,\beta\in\mathbb{R})$, and we take the limit $P\to\infty$ and the $x^{4n}$-direction to periodic direction with the periodicity $\beta$ as $\mathbb{R}^{4n}\to\mathbb{R}^{4n-1}\times S^1$.
This situation replace the well-separated limit ($(a^{4n}_p-a^{4n}_{p+1})^2\gg \lambda^2$ for all $p$) to the condition of the size $\lambda$ and the periodicity $\beta$:
\begin{equation}
\beta\gg\lambda.	\label{eq:small_size_limit}
\end{equation}
In addition, $\phi_{\text{'t Hooft}}(x)$ becomes
\begin{align}
\lim_{p\to\infty}\phi_{\text{'t Hooft}}(x) &= 1+\sum_{p=-\infty}^{\infty}\frac{\lambda^2}{\|\mathbf{x}-\mathbf{a}_x\|^2+\left( t- (a_t-p\beta) \right)^2}	\notag \\
&= 1+ \mu^2\lambda^2\sum_{p=-\infty}^{\infty}\frac{1}{\mu^2\|\mathbf{x}-\mathbf{a}_x\|^2 + \left( \mu(t-a_t)+2\pi p \right)^2},
\end{align}
where $\mu=2\pi/\beta$. Note that we demand the condition $2\pi\gg\mu^2\lambda^2$ from $\beta^2\gg\lambda^2$, but this condition does not have an influence on that we take the factor $\mu^{-2}$ from the dominator.
Now we use the formula:
\begin{equation}
\sum_{p=-\infty}^{\infty}\frac{1}{a^2 + (b+2\pi p)^2} = \frac{\sinh a}{2a(\cosh a - \cos b)},
\end{equation}
then a gauge field of the HS type one-caloron in the higher dimensions ($n\geq2$) is given by
\begin{equation}
A_{\mu} = \frac{1}{4}\Sigma_{\mu\nu}^{(\pm)}\partial_{\nu}\ln\left( 1+\frac{\pi\lambda^2\sinh(2\pi r/\beta)}{\beta r\left( \cosh(2\pi r/\beta) - \cos(2\pi\tilde{t}/\beta) \right)} \right),~~~\text{with}~\beta\gg\lambda.
\end{equation}
Here $r=\sqrt{(x^i-a^i)^2},~\tilde{t}=x^{4n}-a^{4n}$, for any $a^{4n}\in[0,\beta)$ and the index $i=1,\dots,4n-1$.
The condition $\beta\gg\lambda$ means that the caloron's size modulus $\lambda$ is much smaller than the periodic coordinate size $\beta$, hence we call this condition \eqref{eq:small_size_limit} as the small size limit.

In four dimensions, the HS one-caloron becomes the gauge-equivalent to the BPS one-monopole when we take the limit $\beta/2\pi\lambda\to0$ \cite{Rossi:1978qe,Chakrabarti:1987kz}.
On the other hand, in the higher dimensions, the HS type one-caloron requires the small size limit $\beta/\lambda\gg1$, hence the monopole limit $\beta/2\pi\lambda\to0$ is evidently inconsistent with this limit.
Therefore we can not take the monopole limit for the HS type one-caloron in higher dimensions.

\section{Conclusion and discussions}
In this paper, we have studied the ADHM construction of the $4n$-dimensional (anti-)self-dual instantons with the unitary group U$(N)$.
This scheme is the straightforward generalization of the eight-dimensional one \cite{Nakamula:2016srw}.
The (anti-)self-dual basis $e_{\mu}$ which is the generalization of the quaternion basis plays important roles in this scheme.
We have shown that the $4n$-dimensional (anti-)self-dual basis can be produced from the $(4n-1)$-dimensional complex Clifford algebra $C\ell_{4n-1}(\mathbb{C})$.
Moreover, we have explicitly constructed this basis by giving the explicit representation of the complex Clifford algebras.
We have found that there are two ADHM constraints, one of these is the straightforward generalization of four-dimensional one.
The another ADHM constraint, which is the new constraint, corresponds to the non-linearity of the higher-dimensional (anti-)self-dual equation.
One of the most interesting things is that the more non-linearity of the (anti-)self-dual equations is according as dimensions increase but the ADHM construction does not need essentially new constraints more than what is shown in this paper.

We have shown that our construction reproduces the known BPST type one-instantons in higher dimensions.
Furthermore, we have discussed the multi-instantons ADHM ansatz by generalizing the four-dimensional 't Hooft multi-instantons ADHM data.
However, we have found that this ansatz does not satisfy the second ADHM constraint in general and approximately satisfies this constraint only if we take the well-separated limit.
From this fact, we found that the higher-dimensional HS type one-caloron needs the small size limit and can not take the monopole limit to this caloron.
Note that this fact does not mean that the higher dimensional monopoles do not exist.
Indeed, the higher dimensional monopoles are studied in various context \cite{OBrien:1988whk,Radu:2005rf,Tchrakian:2010ar}.
However, only the one-monopoles are (anti-)self-dual in higher dimensions because the first-order equations, which correspond to the generalization of the Bogomol'nyi equation to higher dimensions, in $(4n - 1) \geq 7$ being overdetermined.
Hence, the relation between instantons, monopoles, and calorons in higher dimensions is an interesting topic.

We mention the relation between instantons and Skyrmions.
In four dimensions, Atiyah and Manton pointed out that the holonomy of the Yang-Mills instantons gives a well approximated Skyrmion solutions \cite{Atiyah:1989dq}.
This scheme is known as the Atiyah-Manton construction, although the origin of this approximation was not transparent.
Sutcliffe has shown that a systematic derivation of the energy functional for the Skyrme model from the Yang-Mills action in four dimensions and elucidated the origin of the Atiyah-Manton construction \cite{Sutcliffe:2010et}.
In higher dimensions, we can derive the energy functional for the higher-dimensional Skyrme model from the generalized Yang-Mills action \eqref{eq:gYM_model} by the same method with Sutcliffe.
Since the Derrick's theorem, we show that there are soliton solutions which we call the higher-dimensional Skyrmion in the model.
In the context \cite{Nakamula:2016wwv}, we have found the numerical solution of the above mentioned Skyrmion in eight dimensions.  
Moreover, we have calculated a field through the Atiyah-Manton construction applied to the eight-dimensional 't Hooft type one-instanton and have found that this gave a good approximation to the numerical solution of the Skyrmion.
These results strongly suggest that the instanton/Skyrmion correspondence holds even in $4n$ dimensions and this relation is a universal property.

Finally, many open problems of the higher-dimensional ADHM construction remain, and we list a few of these problems in the following.
\begin{itemize}
\item We have obtained the multi-instantons ADHM data in the well-separated limit, but these moduli parameters were restricted.
Can we construct a strict multi-instantons ADHM data in higher dimensions?
\item In this paper, we have only treated the ADHM data of which the rank of unitary group is $N=2^{2n-1}$.
What are forms ADHM data of which the other ranks of the unitary group?
\item Similarly, can we construct an ADHM construction with other classical groups, for instance, SO$(N)$, Sp$(N)$?
\item In higher dimensions, how many numbers of the moduli parameters does the (anti-)self-dual $k$-instanton have in general?
Moreover, can we show the one to one correspondence between instanton moduli space and ADHM moduli space?
\item In four dimensions, there is the general scheme to construct the monopoles and calorons, known as the Nahm constructions.
Can we construct a Nahm construction in higher dimensions?
\item In four dimensions, the noncommutative ADHM construction is an interesting topic \cite{Nekrasov:1998ss,Nekrasov:2000ih,Hamanaka:2013vca}.
Can we construct a noncommutative ADHM construction in higher dimensions?
\end{itemize}

\subsection*{Acknowledgments}
I am grateful to A. Nakamula and S. Sasaki for many very helpful discussions and comments.
I would like to also thank T. Tchrakian for his valuable comments.
This work is supported by Kitasato University Research Assistant Grant.

\begin{appendix}

\section{Matrix representation of the complex Clifford algebras}
The $m$-dimensional complex Clifford algebra is defined as the algebra that satisfies the relation $\{ \Gamma_i,\Gamma_j \}=\pm2\delta_{ij}~~(i,j=1\dots,m)$.
In this paper, we choose the minus sign of the above relation to constructing the (anti-)self-dual basis.
In the section 2, we have shown that the $4n$-dimensional (anti-)self-dual basis $e_{\mu}$ is constructed from the $(4n-1)$-dimensional complex Clifford algebra $C\ell_{4n-1}(\mathbb{C})$.
Therefore we need the matrix representation of the complex Clifford algebras $C\ell_{4n-1}(\mathbb{C})$ to obtain the explicit representation of (anti-)self-dual basis $e_{\mu}$.
The matrix representation of the complex Clifford algebras is similar as the gamma matrix in the physics, thus we can obtain this matrix representation by using the well known following representation of the gamma matrices:
\begin{align}
\Gamma_1^{(\pm)} &= \pm i\sigma^{(1)}_1\otimes\sigma^{(2)}_3\otimes\dots\otimes\sigma^{(2n-1)}_3, & \Gamma_2^{(\pm)} &= \pm i\sigma^{(1)}_2\otimes\sigma^{(2)}_3\otimes\dots\otimes\sigma^{(2n-1)}_3, \notag \\
\Gamma_3^{(\pm)} &= \pm i\sigma^{(1)}_0\otimes\sigma^{(2)}_1\otimes\dots\otimes\sigma^{(2n-1)}_3, & \Gamma_4^{(\pm)} &= \pm i\sigma^{(1)}_0\otimes\sigma^{(2)}_2\otimes\dots\otimes\sigma^{(2n-1)}_3, \notag \\
&\vdots&&\vdots   \notag \\
\Gamma_{4n-3}^{(\pm)} &= \pm i\sigma^{(1)}_0\otimes\sigma^{(2)}_0\otimes\dots\otimes\sigma^{(2n-1)}_1, & \Gamma_{4n-2}^{(\pm)} &= \pm i\sigma^{(1)}_0\otimes\sigma^{(2)}_0\otimes\dots\otimes\sigma^{(2n-1)}_2, \notag \\
\Gamma_{4n-1}^{(\pm)} &= \pm i\sigma^{(1)}_3\otimes\sigma^{(2)}_3\otimes\dots\otimes\sigma^{(2n-1)}_3, \label{eq:Clifford_algebra_spinor_representation}
\end{align}
where $\sigma_i$ are the Pauli matrices and $\sigma_0=\mathbf{1}_2$.
In addition, we can use the real Clifford algebra $C\ell_{4n-1}(\mathbb{R})$ instead of the complex one to construct the (anti-)self-dual basis $e_{\mu}$.
However, it is difficult in general that we construct the matrix representations of the real Clifford algebras in higher dimensions
\footnote{In the case of $n=1,2$, we can explicit construction the matrix representations of the real Clifford algebra \cite{Nakamula:2016srw}.}.
If we use the real Clifford algebra then the hermite conjugate is replaced the transpose in this paper, and the gauge group of the ADHM instantons becomes the special orthogonal group.
This discussion in more detail in \cite{Nakamula:2016srw}.

We note that the representation of $\Gamma^{(\pm)}_i$ is not uniqueness because there is the following transformation that holds the relation $\{ \Gamma_i^{(\pm)},\Gamma_j^{(\pm)} \} = -2\delta_{\mu\nu}\mathbf{1}_{2^{2n-1}}$:
\begin{equation}
\Gamma_{\mu}^{(\pm)}\mapsto\tilde{\Gamma}^{(\pm)}_i = M\Gamma^{(\pm)}_iM^{-1},	\label{eq:decomposed_Clifford_freedom}
\end{equation}
where $M\in\text{U}(2^{2n-1})$.
Therefore we find that the (anti-)self-dual basis $e_{\mu}$ has the following freedom for the representations:
\begin{equation}
e_{\mu}\mapsto \tilde{e}_{\mu}=Me_{\mu}M^{-1},~~~e_{\mu}^{\dagger}\mapsto \tilde{e}_{\mu}^{\dagger}= Me_{\mu}^{\dagger}M^{-1}.
\end{equation}

\section{The poof that the existence of the inverse $E_k^{(a)}~(a=1,2)$}
We show again the ADHM constraints for convenience.
\begin{align}
\Delta^{\dagger}\Delta = \mathbf{1}_{2^{2n-1}}\otimes E^{(1)}_k,	\\
C^{\dagger}\Delta(\Delta^{\dagger}\Delta)^{-1}\Delta^{\dagger}C = \mathbf{1}_{2^{2n-1}}\otimes E^{(2)}_k.
\end{align}

Lets the Weyl operator satisfy the non-degeneracy condition:
\begin{equation}
\text{rank}~\Delta = 2^{2n-1}k.	\label{eq:Weyl_op_non-degeneracy}
\end{equation}
First, we show the existence of the inverse $E^{(1)}_k$.
Because of the property of the rank: $\text{rank}~\Delta=\text{rank}~\Delta^{\dagger}\Delta=2^{2n-1}k$.
We recall that $\Delta$ is the $(N+2^{2n-1}k)\times2^{2n-1}k$ matrix thus $\dim \Delta^{\dagger}\Delta=2^{2n-1}k$, and we use the rank-nullity theorem: $\dim\Delta^{\dagger}\Delta = \text{rank}~\Delta^{\dagger}\Delta + \text{Ker}~\Delta^{\dagger}\Delta$, then we obtain
\begin{align}
\text{Ker}~\Delta^{\dagger}\Delta = 0,\label{eq:proof_E^1_existence}
\end{align}
where Ker $A$ means the dimension of the kernel of $A$.
If the kernel dimension of the matrix $A$ is zero then the inverse matrix of $A$ is existence, hence there is $(\Delta^{\dagger}\Delta)^{-1}$ for \eqref{eq:proof_E^1_existence}.
This is just the assurance of the existence of the inverse $E^{(1)}_k$.

Next, we  will prove the existence of the inverse $E^{(2)}_k$. The Weyl operator $\Delta$ contains the coordinate parameter $x$, therefore the non-degeneracy condition \eqref{eq:Weyl_op_non-degeneracy} holds for all $x$
\footnote{
Technically speaking, $\Delta(x)$ does not have to satisfy the non-degeneracy condition at the instantons positions.
}.
Because of this fact and $\Delta=C(x\otimes\mathbf{1}_k)+D$, we obtain
\begin{equation}
\text{rank}~\Delta(\infty) = \text{rank}~C(x\otimes\mathbf{1}_k) = 2^{2n-1}k	\label{eq:proof_E^2_existence_1}
\end{equation}
Now we recall that $x\otimes\mathbf{1}_k=x^{\mu}e_{\mu}\otimes\mathbf{1}_k$ is the $2^{2n-1}k \times 2^{2n-1}k$ invertible matrix, since we can give the inverse of $x$ as $x^{-1}=\frac{x^{\mu}}{\|x\|^2}e^{\dagger}_{\mu}$ explicitly.
Hence,
\begin{equation}
\text{rank}~C=\text{rank}~C(x\otimes\mathbf{1}_k)=2^{2n-1}k.	\label{eq:proof_E^2_existence_2}
\end{equation}
We can also obtain the rank of $D$: $\text{rank}~D=\text{rank}~\Delta(0)=2^{2n-1}k$.
These facts give that $\text{rank}~\Delta=2^{2n-1}k \Rightarrow \text{rank}~C=\text{rank}~D=2^{2n-1}k$, and the inverse fact: $\text{rank}~C=\text{rank}~D=2^{2n-1}k \Rightarrow \text{rank}~\Delta=2^{2n-1}k$ is trivial.
Therefore we obtain $\text{rank}~\Delta=2^{2n-1}k \iff \text{rank}~C=\text{rank}~D=2^{2n-1}k$.
We take the $2^{2n-1}k\times2^{2n-1}k$ matrix $C^{\dagger}\Delta$ to the same situations as \eqref{eq:proof_E^2_existence_1} and \eqref{eq:proof_E^2_existence_2}, and we use the expansion $C^{\dagger}\Delta=C^{\dagger}C(x\otimes\mathbf{1}_k)+C^{\dagger}D$ then
\begin{equation}
\text{rank}~C^{\dagger}\Delta(x) = \text{rank}~C^{\dagger}\Delta(\infty) = \text{rank}~C^{\dagger}C(x\otimes\mathbf{1}_k) = \text{rank}~C^{\dagger}C = \text{rank}~C = 2^{2n-1}k.
\end{equation}
From this fact and the rank-nullity theorem, we obtain $\text{Ker}~C^{\dagger}\Delta=\text{Ker}~\Delta^{\dagger}C=0$.
This means that the maps $C^{\dagger}\Delta:\mathbb{C}^{2^{2n-1}k}\to\mathbb{C}^{2^{2n-1}k}$ and $\Delta^{\dagger}C:\mathbb{C}^{2^{2n-1}k}\to\mathbb{C}^{2^{2n-1}k}$ are bijective, and the map $(\Delta^{\dagger}\Delta)^{-1}:\mathbb{C}^{2^{2n-1}k}\to\mathbb{C}^{2^{2n-1}k}$ is also bijective from the above proof.
Therefore, using the bijective map composition, we find that the map $C^{\dagger}\Delta\circ(\Delta^{\dagger}\Delta)^{-1}\circ\Delta^{\dagger}C:\mathbb{C}^{2^{2n-1}k}\to\mathbb{C}^{2^{2n-1}k}$ becomes bijective.
If a map is bijective then the existence of the inverse map, thus the matrix $C^{\dagger}\Delta(\Delta^{\dagger}\Delta)^{-1}\Delta^{\dagger}C$ is invertible.
Therefore we have shown the existence of the inverse $E^{(2)}_k$.

\end{appendix}

\end{document}